\documentclass[12pt,preprint]{aastex}

\shorttitle{Star clusters in M31}
\shortauthors{Narbutis, Vansevi\v{c}ius, Kodaira et al.}

\begin{document}

\title{A Survey of Star Clusters in the M31 Southwest Field. \\
$UBVRI$ Photometry and Multiband Maps}

\author{D.~Narbutis\altaffilmark{1}, V.~Vansevi\v{c}ius\altaffilmark{1},
K.~Kodaira\altaffilmark{2}, A.~Brid\v{z}ius\altaffilmark{1}, and
R.~Stonkut\.{e}\altaffilmark{1}}

\altaffiltext{1}{Institute of Physics, Savanori\c{u} 231, Vilnius
LT-02300, Lithuania, wladas@astro.lt} \altaffiltext{2}{The Graduate
University for Advanced Studies (SOKENDAI), Shonan Village, Hayama,
Kanagawa 240-0193, Japan}

\begin{abstract}
A new survey of star clusters in the southwest field of the M31 disk
based on the high resolution Subaru Suprime-Cam observations is presented.
The $UBVRI$ aperture CCD photometry catalog of 285 objects ($V\lesssim20.5$\,mag;
169 of them identified for the first time) is provided. Each object
is supplemented with multiband color maps presented in the electronic
edition of the {\it Astrophysical Journal Supplement}. Seventy seven
star cluster candidates from the catalog are located in the {\it
Hubble Space Telescope} archive frames.
\end{abstract}


\section{Introduction \label{section:introduction}}
Detailed studies of the star clusters in M31 are essential to
understanding the evolution mechanisms of disk galaxies and of the
cluster population itself. Many observational surveys of this galaxy
have been devoted to globular clusters; see, e.g., \citet{Kim2007}
and references therein. Due to crowding, plausible detection and
analysis of star clusters projected on the disk of M31 became feasible
only with high-resolution imaging; see \citet{Krienke2007} for an
extensive discussion of the problem. However, {\it Hubble Space
Telescope} ($HST$) observations cover only a small part of the M31
disk, and up-to-date cluster population studies \citep[e.g.,][]{Krienke2007}
are based on $HST$ fields scattered over the area of M31.

A homogeneous (in object detection and photometry) survey of star
clusters in the southwest field of the M31 disk was conducted by
\citet{Kodaira2004} (hereafter Paper I) over an area of $\sim$500
arcmin$^{2}$, making use of the high-resolution imaging capability
of the Subaru Suprime-Cam \citep{Miyazaki2002}. In Paper I we presented
a catalog of 49 prominent compact objects and a catalog of 52 emission
objects. Structural parameters of 49 compact and 2 emission objects
brighter than $V\sim19.0$\,mag were derived from the $V$-band Suprime-Cam
image by \citet{Sableviciute2006,Sableviciute2007}, showing that
they are fainter and span a slightly wider half-light radius range
than the M31 clusters studied by \citet{Barmby2002,Barmby2007}. The
$UBVRI$ ($R$ and $I$ bands are in Cousins system) photometry of these
clusters was performed on the Local Group Galaxy Survey (LGGS;
\citealp{Massey2006}) images by \citet{Narbutis2006a}, resulting in
a smaller scatter in color-color diagrams than photometry data of
the same objects taken from the Revised Bologna Catalogue of M31
globular clusters and candidates compiled by \citet{Galleti2004}.

\begin{figure}
\epsscale{0.50}
\plotone{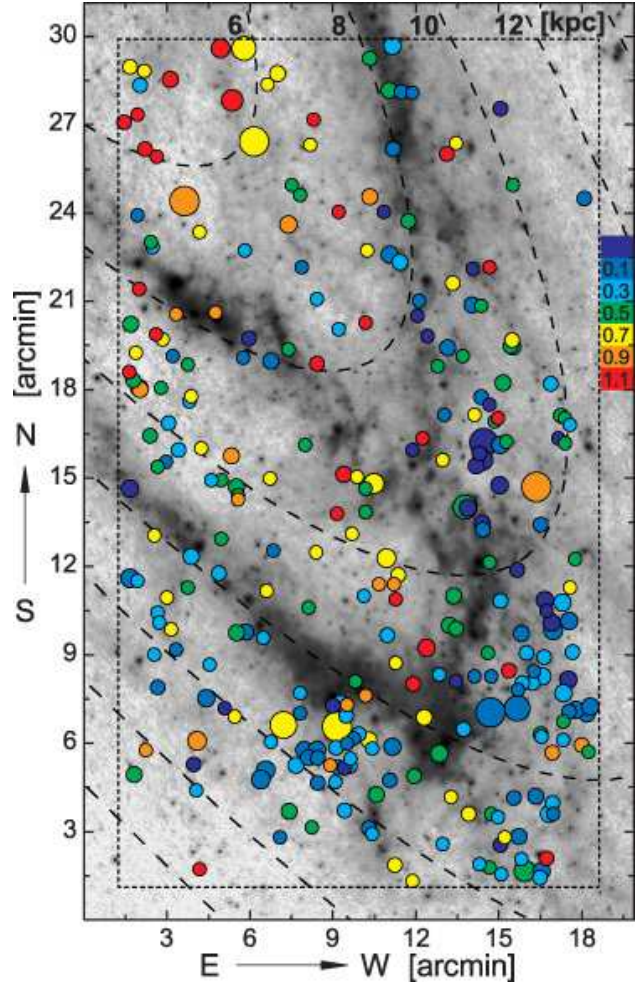}
\caption{Star cluster candidates (285 objects) in the M31 SW field
are overlaid on the {\it Spitzer} (24\,\micron) image. Circle size
corresponds to the $V$-band magnitude, and circle color represents
the observed $B\!-\!V$ (see the color bar for coding). Elliptical
ring segments, indicating distances from the M31 center (6 -- 18\,kpc),
are marked with dashed lines. The rectangle (dotted line) indicates
the Suprime-Cam survey area. North is up, east is left.
\label{figure:field}}
\end{figure}

Promising results from these studies motivated us to extend a sample
of star clusters up to $V\sim20.5$\,mag. Here we present the results
of $UBVRI$ aperture CCD photometry for 285 star cluster candidates
located in the same Suprime-Cam field of the M31 disk (see
Figure~\ref{figure:field}). The multiband color maps, combined from
$GALEX$, LGGS, 2MASS, {\it Spitzer} (24\,\micron), and HI (21\,cm)
images, are provided in the electronic edition of the {\it Supplement}.

We describe the object selection procedure in section
\ref{section:selection}, photometry and calibration in section
\ref{section:photometry}, and present the catalog in section
\ref{section:catalog}.

\section{Object Selection \label{section:selection}}
The Suprime-Cam survey of star cluster-like objects was conducted
in the southwest (SW) field of the M31 disk ($\sim$17\farcm5$\times$28\farcm5
in size), centered at $\alpha_{\rm J2000}=0^{\rm h}40\fm9$ and
$\delta_{\rm J2000}=+40\arcdeg45\arcmin$ (Figure~\ref{figure:field}).
Visual inspection of the high-resolution Suprime-Cam mosaic images,
with characteristic full width at half-maximum (FWHM) of the point-spread
function (PSF) of $\sim$0.7\arcsec, enabled us to select the initial
sample of $\sim$600 star cluster candidates up to $V\sim21^{\rm m}$.

In addition, several objects that were not recognized as cluster
candidates in our survey, were appended to the initial Suprime-Cam
object sample from the Revised Bologna Catalogue of M31 globular clusters
and candidates v.3.2, July 2007\footnote{See http://www.bo.astro.it/M31/.}
\citep{Galleti2007} and from the recent $HST$ survey of the M31 disk
cluster population conducted by \citet{Krienke2007}. About 40 objects,
included in \citet{Galleti2007} and overlapping with our survey field
originally come from the recent survey by \citet{Kim2007}, however,
$\sim$50\% of these objects were omitted from our catalog because
their surface brightness profiles closely resemble that of stars.

To study selection effects inherent to the initial cluster candidate
sample, archival $HST$ frames (see Appendix \ref{appendix:hst}) were
employed. A completeness of $\sim$70\% in the visual selection of
star cluster candidates at the limiting magnitude of the catalogue
($V\sim20.5$\,mag) was estimated by comparing our initial sample of
$\sim$600 objects with an overlapping sample studied by \citet{Krienke2007}.
We also checked our objects against available archival $HST$ images,
and some of our cluster candidates were classified as asterisms.
Moreover, few new star cluster candidates, overlooked during a visual
inspection of the Suprime-Cam images, were found on $HST$ frames.
Summing up results of these tests, we find the completeness of our
cluster candidate sample to be higher than $\sim$50\% at $V = 20.5$\,mag.
This is a conservative magnitude limit for objects that can be identified
as star clusters on the Suprime-Cam images. However, it is difficult
to estimate the completeness of the present cluster candidate sample
more accurately, since it depends on a strongly varying background
object density across the surveyed area, and the clusters' luminosities,
colors, and concentrations.

A strong contamination of young star cluster samples selected in M31
by asterisms has been demonstrated recently by \citet{Cohen2005}.
Therefore, in order to clean our cluster candidate sample ($V\lesssim20.5$\,mag),
we determined cluster structural parameters, by employing the BAOLAB/ISHAPE
program package \citep{Larsen1999} and using the analysis technique
described in \citet{Sableviciute2006}. Unresolved objects with determined
intrinsic sizes of $\lesssim\!0.2$\arcsec\ have been removed as suspected
stars. The analysis of multiband color maps and ISHAPE results revealed
several red objects, possessing smooth elliptical surface brightness
distributions. They were also removed from the sample as probable
background galaxies. We did not include KWC13 and KWC24 from Paper
I either, suspecting them to be background galaxies. After performing
this cleaning, we ended up with the final catalog containing 285 star
cluster candidates (169 of them identified for the first time). The
final cluster candidate sample includes 77 objects observed by $HST$.

\section{Photometry \label{section:photometry}}
For the star cluster aperture photometry, we used
LGGS\footnote{LGGS: http://www.lowell.edu/users/massey/lgsurvey.html.}
$U$, $B$, $V$, $R$, and $I$ band mosaic images of four M31 fields (F6,
F7, F8, and F9) overlapping with the field studied in Paper I. The
mosaic camera used for LGGS consists of eight CCDs. Each CCD chip
covers a 9\arcmin$\times$18\arcmin\ field and has an individual set
of color equations. The observations and data reductions are described
in detail by \citet{Massey2006}.

We considered mosaic images, cleaned of cosmic rays and cosmetic
defects, to be preferable to individual exposure images for star
cluster aperture photometry. The dithering pattern of five individual
exposures is the same for each field (maximum shifts up to 1\arcmin\
from the first exposure), with an exception of the $U$-band mosaic
of the F9 field, which is combined ofrom six individual exposures.
\citet{Massey2006} do not recommend a straightforward use of their
mosaic images for accurate photometry, therefore, we treated each
CCD chip area in the mosaic image separately, taking special care
of objects residing in different CCDs of the combined individual
exposures.

\subsection{PSF homogenization \label{subsection:psf}}
PSFs of the LGGS mosaic images used for aperture photometry differ
significantly (see Table~1). Moreover, four of them have a coordinate-dependent
PSF with a FWHM varying by more than 0.2\arcsec\ across the field.
That would lead to a variable aperture correction and, if not properly
corrected, to a cluster color bias for the small apertures (diameter
of $\sim$3\arcsec) used in this study. Since the intrinsic surface
brightness distribution profiles of star cluster candidates are unknown,
we homogenized mosaic image PSF shapes, instead of using variable
aperture corrections. This also ensures a consistency in aperture
selection, photometric error and photometric background estimates.

We applied the DAOPHOT package \citep{Stetson1987} from the IRAF
program system \citep{Tody1993} to compute original PSFs for all
mosaic images. By convolving the widest PSF (FWHM\,=\,1.3\arcsec)
with the Gaussian kernel, a reference PSF of FWHM\,=\,1\farcs5 was
produced. The IRAF's {\tt psfmatch} procedure was employed to compute
the required convolution kernels for individual mosaic images with
respect to the reference PSF. These kernels were symmetrized by
replacing their cores with the best-fitting Gaussian profiles, and
their wings with the best-fitting exponential profiles, truncated
at 3.5\arcsec. The IRAF's {\tt convolve} procedure was employed to
produce mosaic images possessing unique and coordinate-independent
PSFs of FWHM\,=\,1.5\arcsec. The homogenized images were photometrically
calibrated and used for star cluster photometry. The maximum difference
of the aperture corrections in different pass-bands is less than
0.02\,mag. A test of photometric accuracy of the entire PSF homogenization
procedure suggests that errors do not exceed 0.01\,mag.

\subsection{Calibrations \label{subsection:calibration}}
For the $UBVRI$ cluster photometry calibration we selected well-isolated
stars of high photometric accuracy ($<$0.03\,mag), measured in each
pass-band more than 3 times, from Table~4 of \citet{Massey2006}.
The calibration stars were measured on the homogenized mosaic images
through the circular aperture of 3.0\arcsec\ in diameter by employing
the IRAF's {\tt phot} procedure. The aperture correction (aperture
magnitude minus total magnitude) of 0.27\,mag was determined for
all homogenized mosaic images.

\citet{Massey2006} provide color equations for individual CCD chips
of the mosaic camera (see their Table~2). We solved those equations
by fitting photometric zero-points of all pass-bands for every individual
field and CCD chip. Typically, 80 (ranging from 20 to 140) calibration
stars per chip were used. The final errors of derived zero-points
are less than 0.01\,mag with typical fitting rms\,$<\!0.03$\,mag
for the $I$ band and $<\!0.02$\,mag for other pass-bands. Color
equations given by \citet{Massey2006}, supplemented with the derived
zero-points, were used to transform instrumental magnitudes to the
standard system. For objects located in the mosaic image areas
combined from different CCDs, we used color equations of corresponding
CCDs and performed independent transformations to the standard system.

A comparison of three published stellar photometry data sets in the
SW field of the M31 disk (\citealp{Narbutis2006b}) suggests caution
when using tertiary standards as local photometric standards. However,
a careful reduction, calibration, and internal consistency check
performed by \citet{Massey2006} resulted in millimagnitude differences
between the photometry results of overlapping fields. To our knowledge,
this is the most accurately calibrated photometry data set in the
M31 galaxy to date.

\subsection{Results \label{subsection:results}}
The aperture $UBVRI$ photometry was carried out by employing the
IRAF's XGPHOT package. The images of all objects, except for five
saturated in the $I$ band, are free of visible defects. Apertures
were centered on clusters' luminosity distribution peaks in the
Suprime-Cam $V$-band mosaic image and transformed to individual LGGS
image coordinate systems with the IRAF's {\tt geoxytran} procedure.
In order to minimize cluster photometry contamination by background
objects in crowded fields, we decided to use individual small circular
(elliptical for the two bright objects KW102 and KW141 to avoid
obvious nearby background stars) apertures; their sizes are provided
in Table~2. The photometric background was determined in individually
selected and object-centered circular annuli with typical inner and
outer radii of 3\arcsec\ and 8\arcsec, respectively. For some clusters,
located on a largely variable background, circular background determination
zones were selected individually in representative areas.

\begin{figure}
\epsscale{0.50}
\plotone{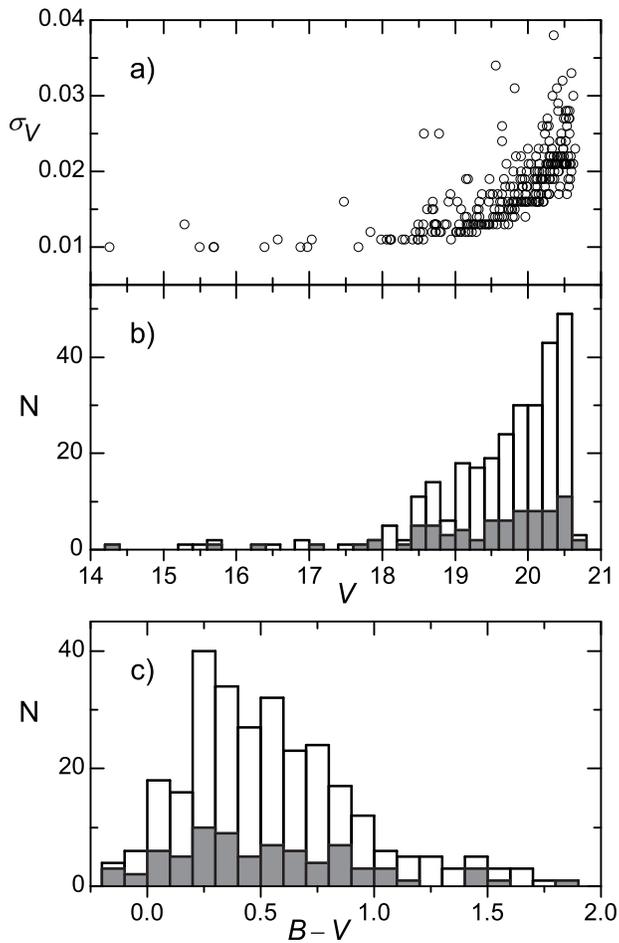}
\caption{Global characteristics of the star cluster candidate catalog
(285 objects). Panels show ({\it a}) the photometric errors, $\sigma_{V}$,
vs. $V$-band magnitudes (we adopted the lowest possible error of
$\sigma_{V}=0.01$\,mag); ({\it b}) the observed $V$-band magnitude
histogram (an object selection criterion is $V\lesssim20.5$\,mag);
({\it c}) the observed $B\!-\!V$ color histogram. Gray-shaded histograms
are constructed for a sub-sample of 77 objects, located in the $HST$
frames.
\label{figure:catalog}}
\end{figure}

The catalog of 285 star cluster candidates consists of 138 and 141
objects measured in two and three different LGGS fields, respectively,
while six objects have been measured in one field. Two main types
of error sources determine the final accuracy of the photometry:
the photon noise -- $\sigma_{\rm n}$ (estimated by XGPHOT) and the
calibration procedure -- $\sigma_{\rm c}$. The $V$-band magnitude
and colors, derived in different LGGS fields for each object, were
examined interactively. Weighted averages were calculated taking
into account individual $\sigma_{\rm n}$ and down-weighting data
derived in the mosaic image areas combined from different CCDs. The
rms of averaged magnitudes and colors characterize calibration errors
in general, therefore, they were assigned to $\sigma_{\rm c}$. The
lowest possible calibration errors of $0.010$, $0.015$, $0.020$\,mag
were set for objects having 3, 2, and 1 independent measurements,
respectively. The final photometric errors, provided in the catalog
(Table~2), are calculated as $\sigma=(\sigma_{n}^2+\sigma_{c}^2)^{1/2}$.
The photometric error values, $\sigma_{V}$, for corresponding $V$-band
magnitudes are plotted in Figure~\ref{figure:catalog}.

In order to check the aperture size effect on the color accuracy and
possible bias due to a contamination by background stars, all objects
were measured through four additional apertures, changing the adopted
size by $\pm$0.6\arcsec\ and $\pm$1.2\arcsec. The results of the
aperture size test, published by \citet{Narbutis2007}, show that
contaminating background stars have the strongest influence in $I$
band. The $U\!-\!B$ color of red objects tends to be systematically
bluer, and an opposite effect is observed for the $V\!-\!I$ of blue
objects. However, in most cases the final photometric errors provided
in the catalog (Table~2) represent the accuracy of cluster colors
well. A comparison of our photometry data with $HST$ observations
of the same objects by \citet{Krienke2007} shows a reasonably good
agreement (\citealp{Narbutis2007}).

\section{The Catalog \label{section:catalog}}
The photometric catalog of 285 star cluster candidates in the SW M31
field is presented in Table~2. Object coordinates, $V$-band aperture
magnitudes, $U\!-\!B$, $B\!-\!V$, $V\!-\!R$, and $R\!-\!I$ colors
with their photometric errors, a flag for 77 objects located in $HST$
frames, and cross-identifications with \citet{Galleti2007}, \citet{Krienke2007},
and Paper I (69, 24, and 58 clusters, respectively) are provided.

All catalog objects, overlaid on the {\it Spitzer} 24\,\micron\ image,
are shown in Figure~\ref{figure:field}. Elliptical ring segments,
indicating distances (6 -- 18\,kpc) from the M31 center in the galaxy's
disk plane, were drawn assuming the following M31 parameters: a distance
modulus of $m-M=24.47$ \citep{McConnachie2005}, center coordinates
$\alpha_{\rm J2000}=0^{\rm h}42^{\rm m}44\fs3$,
$\delta_{\rm J2000}=41\arcdeg16\arcmin09\arcsec$ (NASA Extragalactic
Database), a major axis position angle of 38\arcdeg\ \citep{Vaucouleurs1958},
and a disk inclination angle to the line of sight of 75\arcdeg\
\citep{Gordon2006}.

$V$-band magnitude and $B\!-\!V$ color histograms of the catalog
objects are shown in Figure~\ref{figure:catalog}. Shaded histograms
show a sub-sample of 77 objects identified in the $HST$ frames.

\begin{figure}
\epsscale{0.50}
\plotone{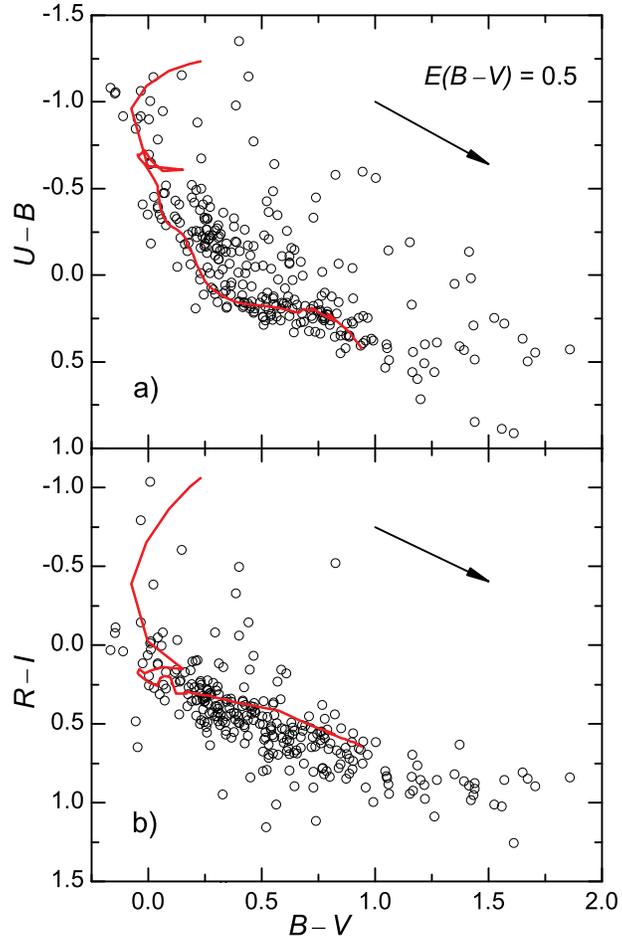}
\caption{Observed color-color diagrams of 285 catalog objects ({\it
circles}). P\'{E}GASE SSP models of Z\,=\,0.008 and ages ranging
from 1 Myr to 15 Gyr are marked with solid lines. The reddening
vectors of $E(B\!-\!V)=0.5$, corresponding to the standard extinction
law, are indicated.
\label{figure:phot}}
\end{figure}

Observed color-color diagrams of star cluster candidates, overplotted
with simple stellar population (SSP) models of metallicity Z\,=\,0.008
and ages ranging from 1 Myr to 15 Gyr, computed with P\'{E}GASE
\citep{Fioc1997}, are presented in Figure~\ref{figure:phot}. Default
P\'{E}GASE parameters and a universal initial mass function
\citep{Kroupa2002} were applied. Reddening vectors are depicted by
applying the standard extinction law: a $V$-band extinction to color
excess ratio $A_{V}/E(B\!-\!V)=3.1$ and color excess ratios
$E(U\!-\!B)/E(B\!-\!V)=0.72$, $E(R\!-\!I)/E(B\!-\!V)=0.69$. The Milky
Way interstellar extinction in the direction of the M31 SW field
is $E(B\!-\!V)=0.062$ \citep{Schlegel1998}.

Figure~\ref{figure:phot}, together with multiband color maps, suggests
that the present sample covers a wide range of stellar populations,
from old globular clusters through young massive clusters. Some objects,
suspected to be young and heavily reddened in the $U\!-\!B$ vs. $B\!-\!V$
diagram, are displaced in opposite $R\!-\!I$ directions from the bulk
of objects in the $R\!-\!I$ vs. $B\!-\!V$ diagram, making a large scatter.
A detailed color analysis of these cases, taking into account multiband
color maps, reveals two main reasons: $R\!-\!I$ is increased due to an
additional flux from red background stars in the $I$-band, and $R\!-\!I$
is decreased by an additional contribution from the $H\alpha$ emission
in the $R$ band.

Combined multiband color maps (Figure Set 4.001-4.285; see Appendix
\ref{appendix:multiband} for description) of 285 objects are provided
in the electronic edition of the {\it Supplement}. They serve as an
illustrative material showing the objects' structure, the location
of the aperture used for photometry, individual background conditions,
and position in the survey field. These maps are also a valuable
tool for revealing cluster interrelations with a global framework
of various M31 galaxy components.

\section{Summary \label{section:summary}}
We have performed the Suprime-Cam survey of star clusters in the
southwest field of the M31 disk up to $V\sim20.5$\,mag, providing
the $UBVRI$ CCD aperture photometry catalog of 285 cluster candidates.
The catalog includes 77 objects located in the $HST$ frames. Photometry
was performed trough individually selected small apertures. Object
cross-identifications with \citet{Galleti2007} and \citet{Krienke2007}
are provided for 69 and 24 objects, respectively. Multiband color
maps combined from LGGS ($U$, $B$, $V$, $I$, and $H\alpha$ bands),
$GALEX$ (NUV, FUV), 2MASS ($J$, $H$, and $K_{\rm s}$ bands), {\it
Spitzer} (24\,\micron), and HI (21\,cm) images, are available in
the electronic edition of the {\it Supplement}.

This catalog contains an almost complete homogeneous sample of target
objects in the surveyed area and will serve as a basis for follow-up
imaging or spectroscopic studies. The presented materials suggest that
the sample in the catalog covers a wide range of stellar population
-- from old globular clusters through young massive clusters. An
analysis of the catalog will be forthcoming (Vansevi\v{c}ius et al.
2009).

\acknowledgments
We are indebted to Ieva \v{S}ablevi\v{c}i\={u}t\.{e} for her help with
the BAOLAB/ISHAPE package. We are thankful to the anonymous referee for
constructive suggestions. This work was financially supported in part
by a Grant of the Lithuanian State Science and Studies Foundation. The
star cluster survey is based on the Suprime-Cam images, collected at
the Subaru Telescope, which is operated by the National Astronomical
Observatory of Japan. The research is based in part on archival data
obtained with the {\it Spitzer Space Telescope}, and has made use of
 the NASA/IPAC Extragalactic Database (NED) and the NASA/IPAC Infrared
Science Archive, which are operated by the Jet Propulsion Laboratory,
California Institute of Technology, under contract with the National
Aeronautics and Space Administration; the SAOImage DS9, developed by
Smithsonian Astrophysical Observatory; and the USNOFS Image and
Catalogue Archive operated by the United States Naval Observatory,
Flagstaff Station. The data presented in this paper were partly obtained
from the Multimission Archive at the Space Telescope Science Institute.

\appendix
\section{$HST$ Frames \label{appendix:hst}}
The Multimission Archive at the Space Telescope Science Institute
(MAST)\footnote{$HST$: http://archive.stsci.edu/hst/search.php.} was
searched for $HST$ frames overlapping with the Suprime-Cam survey
field and publicly available to the date of 7 August 2007. Flat-fielded
frames were cleaned of cosmic rays and corrected for distortions by
employing procedures of the IRAF's STSDAS package. The World Coordinate
System information was corrected for all $HST$ frames by referencing
the Suprime-Cam $V$-band image, which was registered to the USNO-B1.0
catalog system. The numbers of analyzed $HST$ data sets/telescope
pointings overlapping with the studied Suprime-Cam field are: 222/29
(WFPC2), 9/3 (ACS), 43/6 (STIS), and 75/19 (NICMOS). Star cluster
candidates from the catalog, presented in Table~2, have been identified
only in the WFPC2 (71) and the ACS (6) frames -- 77 objects in total.

\section{Multiband Color Maps \label{appendix:multiband}}
The LGGS mosaic images \citep{Massey2006} in the $U$, $B$, $V$, $I$,
and $H\alpha$ bands of the F7 or F8 fields were used for a multiband
map construction. For objects located to the south of the F7 field,
images of lower resolution from field F8 were substituted.

The 2MASS $J$, $H$, and $K_{\rm s}$ band ``1$\times$'' and ``6$\times$''
survey images were retrieved from the NASA/IPAC Infrared Science
Archive\footnote{2MASS: http://irsa.ipac.caltech.edu/.} and co-added
to increase their signal-to-noise ratio. In general, the ``6$\times$''
co-added image was used, except for the gap region in the north of
the surveyed field, where the ``1$\times$'' image was substituted.
The $GALEX$ NUV and FUV images by \citet{GildePaz2006} were retrieved
from the $GALEX$ Atlas of Nearby Galaxies\footnote{$GALEX$:
http://archive.stsci.edu/prepds/galex\_atlas/index.html.} and co-added
to increase a signal-to-noise ratio. The {\it Spitzer} MIPS (24\,\micron)
post-BCD (basic calibrated data) images of M31 (Program ID 99; PI:
G. Rieke) were retrieved from the {\it Spitzer Space Telescope}
Science Center Data Archive\footnote{{\it Spitzer}:
http://ssc.spitzer.caltech.edu/archanaly/.}, and mosaicked using the
SWarp\footnote{See http://terapix.iap.fr/.} package (author E. Bertin).

The HI (21\,cm) image was retrieved from the National Radio Astronomy
Observatory image gallery\footnote{HI: http://www.nrao.edu/imagegallery/php/level3.php?id=475.}
(image courtesy of NRAO/AUI and David Thilker [JHU], Robert Braun
[ASTRON], WSRT). It was converted from the TIFF to the FITS format
using ImageTOOLSca\footnote{See http://arnholm.org/software/index.htm.}.
The sky coordinate grid indicated by \citet{Westmeier2005} in their
Figure~1 on the same HI image, was used for the initial registration.
Fine adjustments were made assuming a global correlation between the
{\it Spitzer} (24\,\micron) and HI (21\,cm) source distributions,
noted by \citet{Gordon2006}.

The Suprime-Cam $V$-band mosaic image coordinate system was used as
a reference to register and transform all images to the homogeneous
pixel scale of 0.2\arcsec\ pixel$^{-1}$. The IRAF's {\tt geotran}
procedure was used for the LGGS, $GALEX$, and 2MASS image transformations;
the {\tt wregister} procedure was used for the {\it Spitzer} (24\,\micron)
and HI (21\,cm) images. Most defects were masked prior to transformations.
Since the LGGS images cover a wide wavelength range, they served as
a reference for registering other images. A visual inspection was
carried out to ensure an accurate coordinate match of the images
prepared for a multiband map production. The FWHM and original pixel
scale values of those images are provided in Table~1.

The sub-images of 80\arcsec$\times$80\arcsec, centered on the objects'
position, were cut out from the transformed images. SAOImage DS9 was
employed to combine them into multiband color maps. The LGGS, 2MASS,
and $GALEX$ sub-images are displayed in a linear intensity scale,
and the {\it Spitzer} sub-images are displayed in a square root scale.
The minimal shown data level is individual for each sub-image, depending
on the background intensity; however, the range of displayed intensities
was kept constant for all objects. The constant displayed intensity
limits across the surveyed area were applied only for HI (21\,cm)
sub-images and are overplotted with contour lines. A gray-shaded
inset shows a global HI emission intensity, with black standing for
the highest and white -- for the lowest signal level over the survey
area. Note, however, that the HI image is converted from the TIFF
format and does not represent a real 21\,cm emission distribution,
serving for illustrative purposes only.

\begin{figure}
\epsscale{1.00}
\plotone{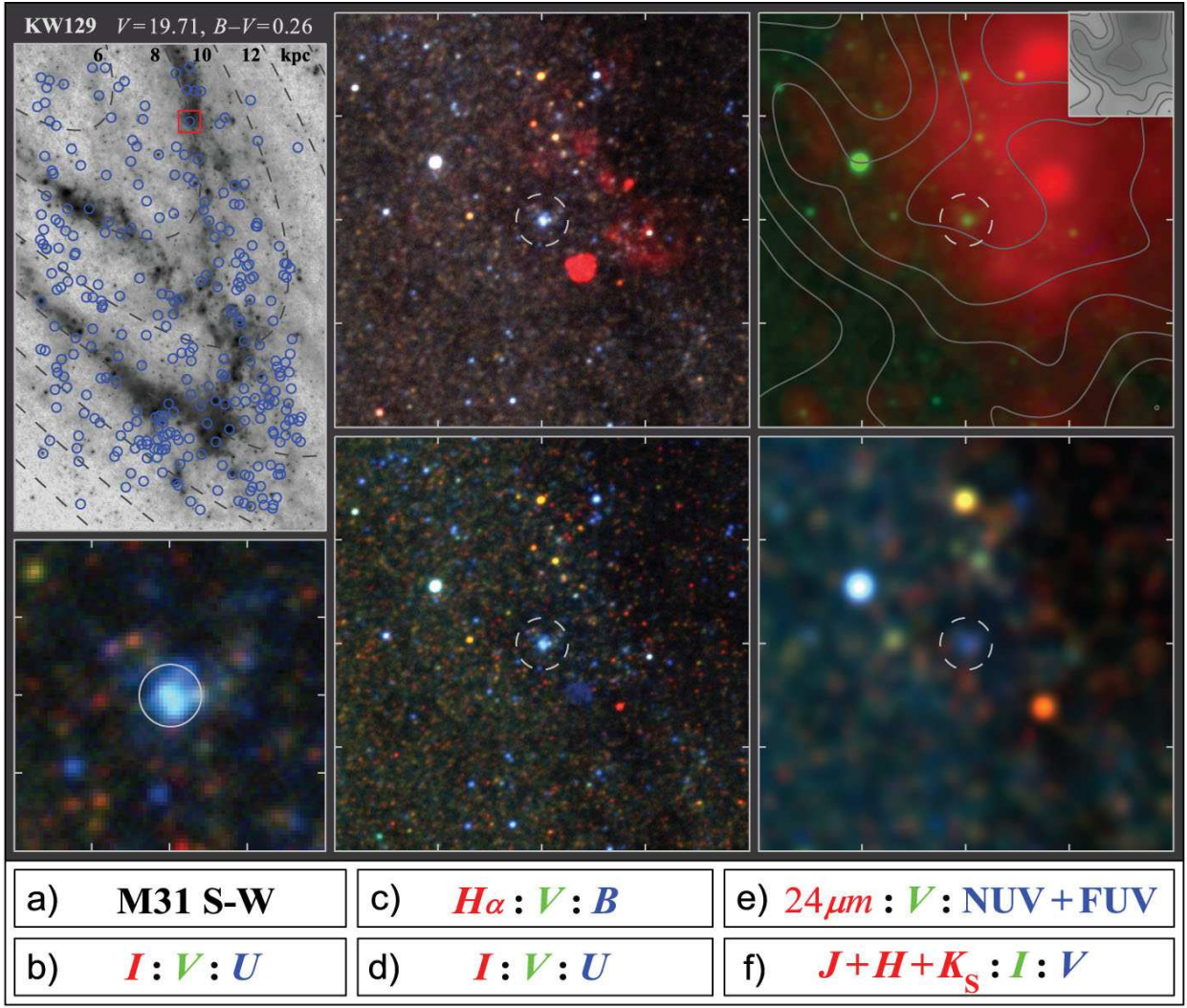}
\caption{\small An example of multiband maps ({\it top}) and a
corresponding layout template ({\it bottom}), indicating panel notations,
is shown here as a guide to the individual multiband color maps,
constructed for all objects, available in the electronic edition of
the {\it Supplement}. The object's ID, the $V$-band magnitude, and
the $B\!-\!V$ color are provided in the top-left corner. Panel {\it
a} shows positions of 285 objects in the M31 SW field ({\it circles});
elliptical ring segments indicating distances from the M31 center
(6 -- 18\,kpc) are marked with dashed lines overlaid on the {\it
Spitzer} (24\,\micron) image; the object under consideration is
indicated by a red square of a size equivalent to the size of panels
{\it c}-{\it f}. In panels {\it b}-{\it f} of the layout template,
passbands of images displayed in red, green, and blue color channels
are indicated by corresponding colors. The LGGS images were used in
panels as follows: ({\it b}-{\it d}) $U$, $B$, $V$, $I$, $H\alpha$
bands of the original resolution; ({\it e}) the $V$-band of a homogenized
PSF (FWHM\,=\,1.5\arcsec); ({\it f}) the $V$ and $I$ bands of a
homogenized PSF (FWHM\,=\,3.5\arcsec) matching the PSF of 2MASS images.
Contour lines in panel {\it e}) emphasize the HI distribution; the
gray-shaded inset shows a global 21\,cm emission intensity, black
standing for the highest and white for the lowest signal level over
the entire surveyed area. The size of panel {\it b} is 15\arcsec$\times$15\arcsec,
and the size of panels {\it c}-{\it f} is 80\arcsec$\times$80\arcsec.
The aperture used for the photometry is overlaid in panel {\it b}.
The object in panels {\it c}-{\it f} is marked with a circle of
10\arcsec\ in diameter. North is up, east is left. [{\it See the
electronic edition of the Supplement for figures 4.1 -- 4.285.}]
\label{figure:example}}
\end{figure}

An example of the online multiband color maps and a layout template
are shown in Figure~\ref{figure:example}. Since the displayed intensity
range is kept constant for all objects, colors of individual objects
are roughly preserved. However, the real emission level at 24\,\micron\
should be estimated by referencing the object's position in the global
{\it Spitzer} image of the survey field. Multiband maps for all 285
objects are provided in a Figure Set 4.001-4.285, published only in
the electronic edition of the {\it Journal}.

\clearpage
\begin{deluxetable}{c c c c} \tabletypesize{\scriptsize}
\tablewidth{0pc} \tablecolumns{4}
\tablecaption{Parameters of Original Images.}
\tablehead{
\colhead{Source} &
\colhead{Images} &
\colhead{FWHM [\arcsec]} &
\colhead{Scale [\arcsec pixel$^{-1}$]}}
\startdata
 Suprime-Cam               & $B$, $V$, $R$, $H\alpha$            & 0.6--0.7                 & 0.20 \\
 LGGS \tablenotemark{a}    & $U$, $B$, $V$, $R$, $I$, $H\alpha$  & 0.7--1.3                 & 0.27 \\
 $GALEX$ \tablenotemark{b} & FUV, NUV                            & $\sim\!4.5$, $\sim\!5.5$ & 1.50 \\
 2MASS                     & $J$, $H$, $K_{\rm s}$               & $\sim\!3.5$              & 1.00 \\
 {\it Spitzer}             & 24\,\micron                         & $\sim\!6.0$              & 2.45 \\
 HI \tablenotemark{c}      & 21\,cm                              & \nodata                  & 10.3 \\
\enddata
\tablenotetext{a}{Mosaic images of the F6--F9 fields.}
\tablenotetext{b}{A strongly variable asymmetric PSF.}
\tablenotetext{c}{A TIFF format image.}
\end{deluxetable}

\clearpage
\begin{deluxetable}{c c c r r r r r r r r r r c c c c c c}
\tabletypesize{\scriptsize}
\setlength{\tabcolsep}{2.0pt} \tablewidth{470pt} \tablecolumns{17}
\tablecaption{$UBVRI$ photometry data of the star cluster candidates in the M31 SW field.}
\tablehead{
\colhead{ID\tablenotemark{a}} &
\colhead{$\alpha_{\rm J2000}$} &
\colhead{$\delta_{\rm J2000}$} &
\colhead{$V$\tablenotemark{b}} &
\colhead{$U\!-\!B$} &
\colhead{$B\!-\!V$} &
\colhead{$V\!-\!R$} &
\colhead{$R\!-\!I$} &
\colhead{$\sigma_{V}$} &
\colhead{$\sigma_{U\!-\!B}$} &
\colhead{$\sigma_{B\!-\!V}$} &
\colhead{$\sigma_{V\!-\!R}$} &
\colhead{$\sigma_{R\!-\!I}$} &
\colhead{$n$\tablenotemark{c}} &
\colhead{Ap\tablenotemark{d}} &
\colhead{$HST$\tablenotemark{e}} &
\colhead{CrossID\tablenotemark{f}}}
\startdata
 KW001 & 10.042834 & 40.607317 &   18.791 &   -0.259 &    0.295 &    0.306 &    0.450 &   0.012 &   0.022 &   0.022 &   0.016 &   0.025 & 3 &  3.2 &       W &           1:BH02 \\
 KW002 & 10.044094 & 40.580988 &   19.963 &    0.197 &    0.636 &    0.473 &    0.524 &   0.016 &   0.027 &   0.025 &   0.022 &   0.028 & 3 &  3.0 &       W &          \nodata \\
 KW003 & 10.045193 & 40.897594 &   19.575 &   -0.256 &    0.211 &    0.241 &    0.353 &   0.013 &   0.035 &   0.022 &   0.018 &   0.022 & 2 &  3.0 & \nodata &          \nodata \\
 KW004 & 10.045684 & 40.603264 &   18.497 &   -0.419 &    0.203 &    0.222 &    0.268 &   0.013 &   0.031 &   0.017 &   0.017 &   0.017 & 3 &  3.0 &       W &      2:37, 3:C01 \\
 KW005 & 10.049319 & 40.585042 &   19.467 &   -0.599 &    0.945 &    0.690 &    0.750 &   0.019 &   0.025 &   0.025 &   0.023 &   0.020 & 3 &  3.0 &       W &          \nodata \\
 KW006 & 10.054298 & 40.605353 &   18.711 &   -0.231 &    0.283 &    0.284 &    0.429 &   0.013 &   0.023 &   0.021 &   0.021 &   0.026 & 3 &  3.4 &       W &          \nodata \\
 KW007 & 10.054684 & 40.691637 &   20.513 &    0.124 &    0.618 &    0.510 &    0.651 &   0.028 &   0.031 &   0.037 &   0.037 &   0.047 & 2 &  3.0 & \nodata &          \nodata \\
 KW008 & 10.056707 & 40.638101 &   19.640 &   -0.282 &    0.379 &    0.354 &    0.426 &   0.016 &   0.020 &   0.022 &   0.022 &   0.032 & 3 &  3.0 &       W &          \nodata \\
\enddata
\tablecomments{R.A. and Decl. (J2000) coordinates of aperture centers
are in the USNO-B1.0 catalogue system (degrees). An entry ``99.999''
is for $R\!-\!I$ and $\sigma_{R\!-\!I}$ of 5 objects saturated in the
$I$-band images.}
\tablenotetext{a}{ID number of the star cluster candidate.}
\tablenotetext{b}{Aperture magnitude (note, it is not a total magnitude).}
\tablenotetext{c}{Number of independent measurements in the LGGS fields F6--F9.}
\tablenotetext{d}{Diameter of circular aperture or major axis length of
elliptical aperture (arc-seconds). Elliptical apertures (ratio of minor
to major axis; position angle of the major axis, calculated counterclockwise
from the North direction) are used for objects KW102 (0.75; 120\arcdeg)
and KW141 (0.70; 30\arcdeg).}
\tablenotetext{e}{Identification of 77 objects in the $HST$ instrument
frames: W -- WFPC2 (71), A -- ACS (6).}
\tablenotetext{f}{Cross identifications ``catalog:id''. (1) -- The
Revised Bologna Catalogue of M31 globular clusters and candidates
v.3.2, July 2007 \citep{Galleti2007} -- 69 objects. KW128 is identified
with V203 and KW129 -- with V202, however, coordinates are offset; KW271
and KW273 are identified as one object -- G099. (2) -- \citet{Krienke2007}
``KHM31'' numbers are given only for 24 objects not listed in (1). KW154
has two identifications: KHM31-79 and KHM31-80. (3) -- \citet{Kodaira2004}
``KWC'' and ``KWE'' numbers are given for 58 objects. KW063 is identified
with KWE09 and KW095 -- with KWE19, however, center coordinates and
apertures are slightly different.}
\end{deluxetable}

\clearpage

\clearpage

\clearpage

\end{document}